\documentclass[12pt]{article}
\pdfoutput=1

\newcommand{\be}{\begin{equation}}
\newcommand{\ee}{\end{equation}}
\newcommand{\bea}{\begin{align}}
\newcommand{\eea}{\end{align}}
\newcommand{\nn}{\nonumber}
\newcommand{\bb}{\mathbb}
\newcommand{\mc}{\mathcal}
\def\({\left(}
\def\){\right)}
\newcommand{\half}{\frac{1}{2}}
\renewcommand{\a}{\alpha}
\renewcommand{\b}{\beta}

\newcommand{\m}{\mu}

\newcommand{\e}{\epsilon}

\newcommand{\C}{{\cal C}}
\newcommand{\ie}{{\it i.e.~}}
\def\eg{{\it e.g.~}}

\usepackage{color}
\usepackage{multirow}
\usepackage{graphicx}
\usepackage{hyperref}
\usepackage{amsmath}
\usepackage{amssymb}
\usepackage{amsfonts}   
\usepackage{latexsym}   
\usepackage{slashed}
\usepackage{yfonts} 
\usepackage{verbatim}
\usepackage{wrapfig}

\begin{document}

\centerline{\large{Emergent space-time and the supersymmetric index}}
%\centerline{\large{Which genus has the largest p-ness?}}
\bigskip
\bigskip
\centerline{Nathan Benjamin$^1$, Shamit Kachru$^1$, Christoph A. Keller$^2$, and Natalie M. Paquette$^1$}
\bigskip
\bigskip
\centerline{$^1$Stanford Institute for Theoretical Physics}
\centerline{Department of Physics, Stanford University, Palo Alto, CA 94305, USA}
\medskip
\centerline{$^2$Department of Mathematics, ETH Zurich}
\centerline{CH-8092 Zurich, Switzerland}
\bigskip
\bigskip
\begin{abstract}
It is of interest to find criteria on a 2d CFT which indicate that it
gives rise to emergent gravity in a macroscopic 3d AdS space via holography.  
Symmetric orbifolds 
%${\rm Sym}^N(X)$ 
in the large $N$ limit have partition functions which are consistent
with an emergent space-time string theory with $L_{\rm string} \sim L_{\rm AdS}$.
For supersymmetric CFTs, the elliptic genus 
%(easily computable at a symmetric orbifold point) 
can serve as a sensitive probe of whether
the SCFT admits a large radius gravity description with $L_{\rm string} \ll L_{\rm AdS}$
after one deforms away from the symmetric orbifold point in moduli space.  We discuss
several classes of constructions whose elliptic genera
strongly hint that gravity with $L_{\rm Planck} \ll L_{\rm string} \ll L_{\rm AdS}$ can emerge
at suitable points in moduli space.

\end{abstract}

\newpage

\tableofcontents

\newpage

\section{Introduction}

A central question in quantum gravity, after the advent of holography, has been ``Which quantum field theories
give rise to emergent gravity?''  While the answer to this question far outruns present understanding, a logical
place to start is with the class of dualities relating 2d CFTs to 3d AdS space-times.  The Brown-Henneaux formula
\cite{BH}
\begin{equation}
c = \frac{3 L_{\rm AdS}}{2G}
\end{equation}
immediately suggests that a weakly curved emergent space-time will require large central charge $c \gg 1$ of the
dual CFT.  However, this is far from a sufficient condition.  For instance, for a variety of reasons, no one expects
$N$ copies of the 2d Ising CFT with $N \gg 1$ to be dual to weakly curved, conventional gravity.

Basic conditions going beyond $c \gg 1$ have indeed been derived in recent literature.  Given a sequence
of CFTs ${\cal C}_N$ of central charges $c_N$ asymptoting to $c \to \infty$ at large $N$, dual to AdS gravity theories with sequentially smaller curvatures, the simple condition that the KK spectrum converge at large $N$ (so that at the $n^{\rm th}$ 
mass level, the number of KK modes has stabilized to a fixed, finite number) is already nontrivial. Simple sequences of theories approaching large $c$, like tensor products of a `seed CFT' $X$, already
fail this test. One way to eliminate the majority of 
states in this example is to take an orbifold,
for example by a series ${\cal G}_N$ of subgroups of
the symmetric group on $N$ objects.
Sequences of this form have a convergent spectrum only if
${\cal G}_N$ asymptotes to an oligomorphic group \cite{HR,Belin:2014fna}.
As symmetric orbifolds constitute the most celebrated and widely studied set of examples of AdS$_3$/CFT$_2$ duality,
results derived in this comparatively simple setting are already useful.

Beyond convergence, one can also derive density of states bounds on the resulting large $c$ CFTs.  By requiring
that the number of CFT states with $\Delta > \Delta_{BH}$, with $\Delta_{BH}$ the dimension corresponding to the
mass of the lightest BTZ black hole, reproduce the Bekenstein-Hawking entropies, Hartman-Keller-Stoica were able
to derive a bound on the low-energy density of states of a holographic CFT \cite{HKS}.  The connection between the high energy
(black hole) states and the low-energy theory comes about via modularity of the torus partition function.
The result, roughly, is that Bekenstein-Hawking entropy is reproduced for CFTs whose low-lying spectrum of states
is sparse enough, growing at most with the Hagedorn density characteristic of string theory.  That is, if the
partition function is of the form
\begin{equation}
Z = \sum_n c_n q^n~,~~q = {\rm exp}(2\pi i \tau)
\end{equation}
then
\begin{equation}
c_n \leq e^{2\pi n},~~n < \frac{c}{12}.
\end{equation}

The most famous examples of AdS$_3$/CFT$_2$, like the D1-D5 system on $K3$, are most easily studied at a symmetric
orbifold point in moduli space.  With $Q_1$ D1's and $Q_5$ D5's, the resulting dual CFT is a sigma model with
target ${\rm Sym}^{Q_1 Q_5}(K3)$.  Intuitively, we expect this to describe a point far from the large radius
gravity solution on $AdS_3 \times S^3 \times K3$, and this is correct. Inspired by the success of this
system, it is natural to consider variations of it. Instead of taking the symmetric orbifold of $K3$ or $T^4$,
one can take an arbitrary seed theory $\C$. It turns out that to leading order, the spectrum is completely universal,
\ie does not depend on the details of $\C$, and is given by
\be
c_n \sim \left\{ \begin{array}{ccc} e^{2\pi n} &:& c \ll n \ll cN/6 \\ e^{2\pi\sqrt{cN n/6}} &:&  cN/6\ll n \end{array} \right.
\ee
More generally, one can consider permutation orbifolds, that is orbifolds by subgroups of $S_N$.
The situation is more complicated, but the results suggest that the spectrum is again
more or less independent of the details of $\C$.
 In fact  \cite{BKM} established a universal lower bound
\begin{equation}
c_n \sim e^{\frac{2\pi n}{\log n}}
\end{equation}
for $n$ below the black hole bound, at sufficiently large $N$ for arbitrary $\C$.
In summary, the spectrum at the orbifold point is always universal, to 
leading order independent
of the seed theory $\C$, and grows
faster than supergravity.

The difficult point in finding theories dual to conventional Einstein gravity coupled to a finite tower of low-energy
fields -- i.e., a system with $L_{\rm Planck} \ll L_{\rm string} \ll L_{\rm AdS}$ -- is that the dual CFTs will
certainly not live at simple points in their moduli space, like symmetric orbifold points.  (This is a specific
instantiation of the much more general principle of `conservation of misery' which finds broad applicability in physics and life.)  It would be nice to find a criterion that could detect, already at the symmetric orbifold point, the
presence of a sparse spectrum -- indicative of supergravity as opposed to low-tension string theory growth in the
low-energy density of states.  Such a criterion for 2d SCFTs (with ${\cal N}=(2,2)$ supersymmetry) was proposed by Benjamin-Cheng-Kachru-Moore-Paquette \cite{BCKMP}.
The elliptic genus of a SCFT is constant on the moduli space, and therefore gives a lower bound
for the number of states at
each mass level valid at any point in moduli space.

\begin{figure}
\label{modulispace}
\begin{center}
\includegraphics[width=0.45\textwidth]{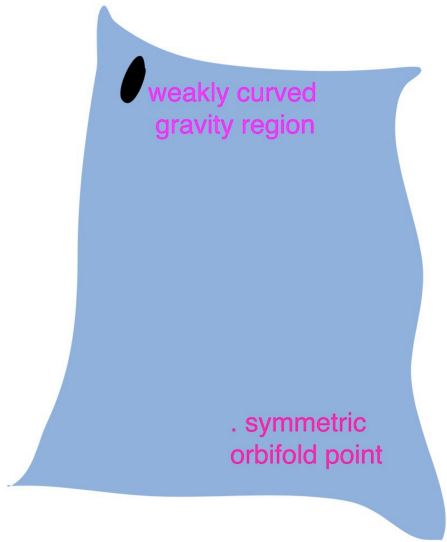}
\end{center}
\caption{The moduli space of an SCFT, with the symmetric orbifold point and the small region dual to large radius gravity labeled.  The elliptic genus computed at any point in the moduli space will match the sparsest spectrum arising anywhere, and so is sensitive to existence of a weakly  curved gravity region.
}

%\label{fig:sym10K3}
\end{figure}

The criterion proposed in \cite{BCKMP} was that if the coefficients in the NS sector elliptic genus exhibit a growth
\begin{equation}
\label{fatrabbit}
c_n \sim {\rm exp}({2\pi n^{p}})
\end{equation}
for any $p < 1$, then the theory should admit a point in moduli space where the dual has $L_{\rm string} \ll L_{\rm AdS}$ (subject to a non-cancellation hypothesis discussed in detail in that reference).
The D1-D5 system on K3 was checked and found to satisfy this condition.

In fact we will find that for the elliptic genus the result does depend on the seed theory $\C$.
In particular, it singles out the $K3$ case $c=6$. This is not altogether surprising.
For a supergravity point to exist in the first place, the theory has to have an
interesting moduli space. Such a modulus has to come from the twisted sector,
since untwisted sector moduli only change the seed theory, but do not move
away from the orbifold point. The weight of the ground state in a twisted
sector of length $n$ is
\be
\Delta= \frac{c}{24}\left(n-\frac{1}{n}\right)\  .
\ee
If $c=12$ or higher, then it is impossible to obtain twist fields with dimension
$\frac{1}{2}$, which would be needed for them to be moduli of the $\mc N=(2,2)$
theory.

A natural question is: can we give general other constraints on (or examples of) constructions which will satisfy
(\ref{fatrabbit})?  That is the burden that this note undertakes.  In \S\ref{sec:toro}, we review salient aspects of weak Jacobi
forms (which arise as elliptic genera of $(2,2)$ SCFTs), and of the Dijkgraaf-Moore-Verlinde-Verlinde formula
determining the genus of a symmetric product in terms of that of the seed CFT.  In \S\ref{sec:miso}, we describe examples of
seed Jacobi forms which, processed via the DMVV formula, satisfy our slow-growth condition.  In \S\ref{sec:wasabi}, we discuss
a specific class of constructions (labelled by a single integer $a$) which yields sequences of CFTs running off to
infinite central charge with a spectrum (in the elliptic genus) consistent with an interpretation involving
supergravity modes up to a string scale $M_{\rm string} \ll M_{\rm Planck}$, at which point the spectrum turns
Hagedorn.  Finally, we conclude in \S\ref{sec:jiro}\ with thoughts about future directions.  Some detailed derivations are relegated to appendices.

\section{Modular kindergarten}
\label{sec:toro}

In this section, we discuss some basic facts about elliptic genera of 2d SCFTs that we will use throughout the
rest of the paper.  We will focus on theories with $(2,2)$ supersymmetry and central charge $c = 6M$ (with integral index $M$) for ease of presentation and to keep the technicalities to a minimum.  Many of our considerations could be extended 
to theories with half-integral index, or with only $(0,2)$ supersymmetry.  

\subsection{The space of weak Jacobi forms}

The elliptic genus of a $(2,2)$ SCFT is defined as
\begin{equation}
Z_{EG}(\tau,z) = {\rm Tr}_{RR}\left( (-1)^{J_0} e^{2\pi z J_0} q^{L_0} (-1)^{\bar J_0} \bar q^{\bar L_0}\right)~.
\end{equation}
In a unitary theory with discrete spectrum and spectral flow symmetry, it is a weak Jacobi form (and in particular, it is a holomorphic object -- all dependence on ${\bar q}$ cancels in the trace) \cite{Kawai:1993jk}.  
For a thorough introduction to weak Jacobi forms, see \eg \cite{Eichler}.  A weak
Jacobi form of \emph{weight} $w$ and \emph{index} $M$ satisfies the modular and elliptic transformation laws
\begin{equation}
\phi \left(\frac{a\tau+b}{c\tau+d}, \frac{z}{c\tau+d}\right) = (c\tau + d)^w e^{2\pi i M \frac{cz^2}{c\tau+d}}\phi(\tau,z)
\end{equation}
and
\begin{equation}
\phi(\tau, z + \ell \tau + \ell^\prime) = e^{-2\pi i M (\ell^2 \tau + 2 \ell z)} \phi(\tau,z), ~~\ell, \ell^\prime \in \mathbb{Z}~~.
\end{equation}
The elliptic genera we will study
have vanishing weight, and their index is 
related to the central charge of the CFT by $c=6M$.
Weak Jacobi forms have Fourier expansions
\be
\phi(\tau,z)=\sum_{m\geq0,\ell}c(m,\ell)q^m y^\ell \qquad q = e^{2\pi i\tau}\ , \ y = e^{2\pi i z}\ .
\ee
We call the \emph{polarity} of a term $4Mm-\ell^2$, and we have
\be
c(m,\ell)=0 \qquad \mathrm{if} \qquad 4Mm-\ell^2 < -M^2\ .
\ee
It is an important fact that weak Jacobi terms are determined
by the terms of negative polarity, which implies that
for a given weight and index the space of such objects is finite dimensional.
In fact, the space of weak Jacobi forms of even weight and integral index is a bi-graded ring
with four generators.  Two of these are the Eisenstein series $E_4$
\begin{equation}
E_4(\tau) = 1 + 240 \sum_{n=1}^\infty \sigma_3(n) q^n
\end{equation}
and $E_6$
\begin{equation}
E_6(\tau) = 1 - 504 \sum_{n=1}^\infty \sigma_5(n) q^n~,
\end{equation}
with index 0.  The others are the forms
\begin{equation}
\phi_{0,1}(\tau,z) = 4\left(\frac{\theta_2(\tau,z)^2}{\theta_2(\tau,0)^2} + \frac{\theta_3(\tau,z)^2}{\theta_3(\tau,0)^2} + \frac{\theta_4(\tau,z)^2}{\theta_4(\tau,0)^2}\right),
\end{equation}
and
\begin{equation}
\phi_{-2,1}(\tau,z) = -\frac{\theta_1(\tau,z)^2}{\eta(\tau)^6}
\end{equation}
with $(w,M)=(0,1)$ and $(-2,1)$ respectively.
Given these generators, it is a simple matter to write down a basis of weight 0 and index $M$ for any desired $M$.  

%For instance, at $m=1$ there is only $\phi_{0,1}$; at index $m=2$, one encounters $\phi_{0,1}^2$
%as well as $E_4 \phi_{-2,1}^2$; and at index $m=3$, generators are $\phi_{0,1}^3$, $E_4 \phi_{0,1}\phi_{-2,1}^2$,
%and $E_6 \phi_{-2,1}^3$.

\subsection{NS sector genus}

Our basic criteria for supergravity growth are most
easily understood in the NS sector.  
Starting from the R sector elliptic genus $\phi(\tau,z)$
of index $M$, we can define by spectral flow the NS sector elliptic genus as
\begin{equation}
\chi(\tau,z) := {\rm exp}\left(2\pi i M \(\frac{\tau}4 + z + \half\) \right) \phi\(\tau, z + \frac{\tau}2 + \half\).
\end{equation}
This object has the transformation properties
\begin{eqnarray}
\chi(-1/\tau,z/\tau) &=& (-1)^M {\rm exp} (2\pi i M z^2/\tau) \chi(\tau,z)\nn\\
\chi(\tau+2,z) &=& (-1)^M \chi(\tau,z)~.
\end{eqnarray}
We will be mostly interested in the specialization
$\chi(\tau):=\chi(\tau,0)$.
The $\chi(\tau)$ is then invariant under a 
subgroup of $\Gamma = SL(2,\mathbb{Z})$.  Namely let $S$ and $T$
denote
the standard generators of $\Gamma$
\begin{equation}
S = \left( \begin{array}{cc}0&-1\\1&0\end{array}\right),~~
T = \left( \begin{array}{cc}1&1\\0&1\end{array}\right)~.
\end{equation}
Then the congruence subgroup $\tilde \Gamma_\theta$ is
defined as the subgroup of $\Gamma$ $\tilde \Gamma_{\theta} \equiv \langle T^4, ST^2 \rangle$.
We see that the NS sector elliptic genus is invariant
under $\tilde \Gamma_\theta$.

It will be useful in the following to know that $\tilde \Gamma_\theta$ is in fact a genus zero subgroup of $SL(2,\mathbb{Z})$, and its Hauptmodul is given by
\begin{equation}
\kappa(\tau) = \left( \frac{2\theta_4}{\theta_2}\right)^2
- \left( \frac{2\theta_2}{\theta_4}\right)^2~=~q^{-1/4} - 20 q^{1/4} - 62 q^{3/4} + \cdots
\label{eq:california}
\end{equation}
Therefore, any NS elliptic genus can be expressed as a 
polynomial in $\kappa$. The degree of this polynomial
is given by $M$, and it is an even or odd polynomial 
if $M$ is even or odd respectively.
Note in particular that
\be\label{kappatrafo}
\kappa(-1/\tau)=-\kappa(\tau)\ , \qquad \kappa(\tau+2)=-\kappa(\tau)\ ,
\ee
so that $\chi(\tau)$ picks up similar signs under $S$
and $T^2$ transformations.

\subsection{The DMVV formula}
In what follows we will need to know the elliptic
genus of symmetric orbifolds.
There is a beautiful formula for the generating
function of symmetric orbifold elliptic genera derived by
Dijkgraaf, Moore, Verlinde and Verlinde \cite{DMVV}.  
If theory $X$ has elliptic genus
\begin{equation}
Z_{EG}^X(\tau,z) = \sum_{m,\ell} c(m,\ell)q^m y^\ell\ ,
\end{equation}
then the generating function
\begin{equation}
{\cal Z}^X(p,\tau,z) = \sum_{N \geq 0} p^N Z^{{\rm Sym}^N(X)}_{EG}(\tau,z) 
\end{equation}
is given by
\begin{equation}
\label{dmvv}
{\cal Z}^X(p,\tau,z) = \prod_{n>0, m, \ell} ~{1 \over {(1 - p^n q^m y^\ell)^{c(nm,\ell)}}}\ .
\end{equation}
That is, there is a simple expression for the generating
functional of the elliptic genera of symmetric products,
in terms of the coefficients $c(m,\ell)$ of the seed elliptic genus.

There is an alternative way of stating this formula in
terms of Hecke operators, which in particular makes
manifest the modular transformation properties of $Z^{{\rm Sym}^N(X)}_{EG}$.  
Recall that the Hecke 
operators $T_L$ act on the elliptic genus via
\begin{equation}
\label{Hecke}
T_L \phi(\tau,z) = \sum_{\substack{ad = L\\b ~{\rm mod}~d}} ~~{1\over L} \phi({{a\tau + b} \over d},az) =
\sum_{ad=L} {1\over a} \sum_{m\geq0,\ell} c(md,\ell) q^{am}y^{a\ell}~.
\end{equation}
In terms of the Hecke operators, the generating functional ${\cal Z}$ takes the simple and elegant form
\begin{equation}\label{DMVVHecke}
{\cal Z}(p,\tau,z) = {\rm exp}\left( \sum_L p^L T_L\phi(\tau,z) \right)~.
\end{equation}

\section{Symmetric orbifolds with slow growth}
\label{sec:miso}

In this section, we show that combining knowledge of the DMVV formula with facts about the growth properties of
the coefficients of basis Jacobi forms at each value of the index, we can find strong restrictions on the
elliptic genera of a seed $X$ if ${\rm Sym}^N(X)$ is to have sub-Hagedorn growth.

\subsection{Basic derivation}
\label{sec:ramen}

In this section, we consider symmetric products of seed (2,2) theory $X$.  Define
\begin{equation}
M = \frac{c_X}{6}~.
\end{equation}
We are interested in the NS sector elliptic genus of the
symmetric orbifold of this theory.
Unless the vacuum contribution cancels somehow, 
$\chi^X$ will start out as $q^{-M/4}$. It is thus
natural to multiply by $q^{M/4}$ to put the leading term in the
NS sector elliptic genus at ${\cal O}(q^0)$. Using
spectral flow we find
from the DMVV formula (\ref{dmvv}) that
\begin{eqnarray}
\sum_{N=1}^\infty p^N Z^{{\rm Sym}^N(X)}_{NS}(\tau,z) &=&
\label{eq:philadelphia}
\sum_{N=1}^\infty p^N Z^{{\rm Sym}^N(X)}_{EG,R}(q,y\sqrt{q}) y^{NM} q^{NM/2}\\
&=&\prod_{\substack{n\geq1, n \in \mathbb{Z}\\ m\geq0, m \in \mathbb{Z}\\ \ell \in \mathbb{Z}}} ~{1 \over {(1- p^n q^{m+ \ell/2 + Mn/2}y^{\ell + Mn})^{c(nm,\ell)}}}~ \nonumber.
\end{eqnarray}
Relabeling indices for convenience as
\begin{eqnarray}
m^\prime &=& m + {\ell \over 2} + {Mn \over 2}\nn\\
\ell^\prime &=& \ell + Mn ~
\end{eqnarray}
and dropping primes gives
\begin{equation}
\label{bigpigpeaches}
\sum_{N=1}^\infty p^N Z_{NS}^{{\rm Sym}^N(X)}(\tau,z) = \prod_{\substack{n\geq1,~n\in\mathbb{Z}\\m\geq\frac{\ell}2,~2m\in\mathbb{Z}\\\ell \in\mathbb{Z},~m-\frac{\ell}2\in\mathbb{Z}}}~~{1 \over {(1- p^n q^m y^\ell)^{c(nm-{n\ell \over 2}, \ell-Mn)}}}~.
\end{equation}

We would like to determine the coefficient of $p^N q^x$ with $x \ll N$ in 
(\ref{bigpigpeaches}), setting $y=1$.  This determines the contribution from CFT
states with $\Delta = x$ in the NS sector. A term that will always be able to
contribute, regardless of the power of $q$, is
\begin{equation}
\label{basicterm}
\frac{1}{(1-p)^{c(0,-M)}}
\end{equation}
assuming $c(0, -M)$ does not vanish.
No other term with $m=0$ can contribute, since the relevant $c$ exponent in the denominator of 
(\ref{bigpigpeaches}) would vanish. The remaining terms in (\ref{bigpigpeaches}) must contribute $q^x$. Suppose some combination of terms contributes $p^a q^x$. Note that $a$ cannot be arbitrarily large due to nonvanishing of $c(nm-\frac{n\ell}2,\ell-Mn)$'s. Thus, in the large $N$ limit, we can take the remaining $p^{N-a}$ from (\ref{basicterm}), and to leading order in $N$, we can ignore the other contributions to $p$. 

Using the Taylor expansion of (\ref{basicterm}), we see that the coefficient of $q^x p^N$ with $x \ll N$ is
\begin{equation}
\label{coefestimate}
{N^{c(0,-M)-1} \over (c(0,-M)-1)!} h(x) + \mc{O}(N^{c(0,-M)-2})
\end{equation}
where $h(x)$ is the coefficient of $q^x$ in the expansion of
\begin{equation}
\label{hungryhippo}
\prod_{\substack{m>0 \\ 2m \in \mathbb{Z}}} ~{1\over (1-q^m)^{f(m)}}~.
\end{equation}
Here $f(m)$ is given by
\begin{equation}
\label{fpig}
f(m):= \sum_{n=1}^\infty \sum_{\substack{\ell \leq 2m \\ \ell \equiv 2m ~({\rm mod} ~2)}}~
c(nm-{n\ell\over 2},\ell - Mn)~.
\end{equation}
Because $c(n,\ell)$ vanishes for $4Mn - \ell^2 < -M^2$, we can write this as
\begin{equation}
\label{fhog}
f(m) = \sum_{n=1}^{\lceil {4m \over M} \rceil} \sum_{\substack{\ell = -2m -M\\ \ell \equiv 2m ~{(\rm mod} ~2)}}^{2m} ~c(nm-{n\ell \over 2}, \ell - Mn)~.
\end{equation}

\subsection{Example: ${\rm Sym}^N(K3)$}
\label{sec:sushi}

We can use the equations (\ref{coefestimate}), (\ref{hungryhippo}), and (\ref{fhog}) to determine
the growth rate of the coefficients rather efficiently.
Let us first illustrate this with the case of the 
symmetric product K3 CFTs.

Since the K3 seed gives an elliptic genus
which has index $M=1$, (\ref{fhog}) becomes
\begin{equation}
f(m) = \sum_{n=1}^{4m} \sum_{\substack{\ell=-2m \\ \ell \equiv 2m ~{(\rm mod}~2)}}^{2m} ~c(nm - {n\ell \over 2}, \ell - n)~.
\end{equation}
We prove
in Appendix \ref{app:soysauce} that in fact, in this case
\begin{equation}
\label{K3f}
f(m) = 
\begin{cases} 
28,~~m \in \mathbb{Z}\\
44,~~m \in \mathbb{Z} + \half.
\end{cases}
\end{equation}

Then following the reasoning of \S\ref{sec:ramen}, and using the fact that $c(0,-1) = 2$ for K3, the growth of the $q^x$ term
in ${\rm Sym}^N(K3)$ (for $x \ll N$) goes as
\begin{equation}
c_{{\rm Sym}^{N}(K3)}(x) = N h(x) + {\cal O}(1)~,
\end{equation}
where $h(x)$ is the coefficient of $q^x$ in 
\begin{equation}
\label{bobloblaw}
\prod_{m=1}^\infty \left( {1\over (1-q^m)^{28}}
{1\over (1-q^{m-{1\over 2}})^{44}}\right)~.
\end{equation}

Now, we are in business.  As far as asymptotics
are concerned, for $1 \ll x \ll N$, (\ref{bobloblaw})
is equivalent to the partition function of 72 free bosons
(up to subleading corrections in powers of $x$).  So
the asymptotic growth at large $x$ is
\begin{equation}
h(x) \sim {\rm exp}(\sqrt{48}\pi\sqrt{x})
\end{equation}
and hence
\begin{equation}
c_{{\rm Sym}^N(K3)}(x) \sim N {\rm exp}(\sqrt{48}\pi \sqrt{x})~.
\end{equation}
This gives a sharper result than the one in \cite{BCKMP}, and indeed has parametrically slower
than Hagedorn growth for states of $\Delta \ll c$.

For the symmetric orbifold of K3, the elliptic genus was analyzed 
carefully in \cite{deBoer:1998ip,deBoer:1998us} and matched to the expected
spectrum of the supergravity states.\footnote{A more general discussion of the
matching between elliptic genera and supergravity BPS spectra in AdS$_3$/CFT$_2$ can
be found in \cite{KL}.} Naively one might have
expected a growth as $\sim e^{x^{5/6}}$ from the modes
coming from $AdS_3\times S^3$. We can obtain a slightly less naive count by considering the growth of quarter-BPS states, as opposed to any arbitrary states. This would predict a growth of $\sim e^{x^{3/4}}$, coming from two constraints: imposing that the right-movers are chiral primary (linking the right-moving weight and $U(1)$ charge), and imposing that the supergravity modes have spin at most two (linking the left and right $U(1)$ charges). For the elliptic genus
however there are sufficiently many cancellations amongst the BPS states that
reduce this growth to $\sim e^{\sqrt{x}}$.
Note that the behavior of the elliptic genus
is drastically different from
the growth of the ordinary partition function,
which exhibits Hagedorn growth \cite{Keller}. There 
are enough cancellations to change the behavior.

\subsection{Results for $f(m)$ for other seed Jacobi forms}
In the previous section we discussed the case $M=1$. Let us
now investigate more general theories.
We found that the properties of $f(m)$ 
determine the growth of the symmetric product elliptic genus.
If $f(m)$ is a constant, one will obtain $e^{\sqrt{x}}$ type growth
for states of energy $x$ far below the Planck scale.  
On the other hand if $f(m) \sim m^\alpha$, one can write (\ref{hungryhippo}) with $q\rightarrow 1-\e$ as
\begin{align}
\prod_{m} \frac{1}{(1-q^m)^{m^\a}} &= e^{-\sum_{m} m^\a \log{(1-q^{m})}} \nn \\
&= e^{\sum_{m}\sum_{n=1}^{\infty} \frac{m^\a}n q^{mn}} \nn \\
&\sim e^{\sum_{n=1}^{\infty} \frac1n \frac{1}{(1-q^n)^{\a+1}}} \nn \\
&\sim e^{\sum_{n=1}^{\infty} \frac1n \frac{1}{(n\epsilon)^{\a+1}}} \nn \\
&\sim e^{\frac{1}{(\log{q})^{\a+1}}}.
\label{eq:horseradish}
\end{align}
We estimate the $q^x$ coefficient of (\ref{eq:horseradish}) by 
\be
c_N(x) \sim \frac{1}{2\pi i} \oint dq ~ e^{\frac{1}{(\log{q})^{\a+1}}+x\log{q}}
\label{eq:seaweed}
\ee
which has a saddle at 
\be
\log q \sim x^{-\frac{1}{\a+2}}.
\ee
When plugged back into (\ref{eq:seaweed}) this finally gives a degeneracy of
\begin{equation}
\label{pness}
c_N(x) \sim {\rm exp}(x^{\frac{\alpha+1}{\alpha+2}}),~~1 \ll x \ll N~.
\end{equation}
Exponential growth of $f(m)$ with $m$ likewise
translates into Hagedorn (or faster) growth of the density of
states, for states far below the energy of the lightest black holes.

An important point about using $f(m)$ is that, as is clear from
(\ref{fpig}) and (\ref{fhog}), it is ${\bf linear}$ in the elliptic
genus of the seed theory.  This means that for a seed of given index
$M$, we can determine all possible behaviors by determining the $f$'s
associated with a basis of the Jacobi forms of weight 0 and index $M$.
Table \ref{table:fm} illustrates the growth of $f$ for the simplest, low
index seed weak Jacobi forms (bases for the seed forms up through $M=4$).

\begin{table}
\resizebox{15cm}{!}{
\begin{tabular}{|c|c|c|c|c|c|c|c|c|c|c|}
\hline Seed & $f(\frac12)$ & $f(1)$ & $f(\frac32)$ & $f(2)$ & $f(\frac52)$ & $f(3)$ & $f(\frac72)$ & $f(4)$ & $f(\frac92)$ & $f(5)$ \\
\hline $2\phi_{0,1}$ & 44 & 28 & 44 & 28 & 44 & 28 & 44 & 28 & 44 & 28 \\
\hline $\phi_{0,1}^2$ & 40 & 380& -4056& 71660& -1327064& 25594236& -507617240 & 10278253676 & -211419320280 & 4403151249788\\
\hline $\phi_{-2,1}^2E_4$ &-8& 284& -4104& 71564& -1327112& 25594140& -507617288& 10278253580&-211419320328& 4403151249692\\
\hline $\phi_{0,1}^3$ &60& 1014& 1816& -14294& -644356& 17951660& -147123588& -2426621198 & 83845804120 & -729784346858 \\
\hline $\phi_{0,1} \phi_{-2,1}^2 E_4$ &12 & 54 & 4120 & -30806 & -561460 & 17523212 & -144884148 & -2438633486 & 83911341400 & -730146191018\\
\hline $\phi_{-2,1}^3 E_6$& -12 & -426 & 5704 & -39062 & -520012 & 17309420 & -143764428 & -2444639630 & 83944110472 & -730327113098\\
\hline $\phi_{0,1}^4$ &80 & 2152 & 17984 & -52844 & -594784 & 21495912 & -541151200 & 11318888452 & -213084244240 & 4272269158504\\
\hline $\phi_{0,1}^2 \phi_{-2,1}^2 E_4$& 32 & 232 & 4736 & 33556 & -1034176 & 21493992 & -521030848 & 11119909252 & -212085289888 & 4272269156584\\ 
\hline $\phi_{0,1} \phi_{-2,1}^3 E_6$& 8 & -728 & -1888 & 81940 & -1253872 & 21493032 & -510970672 & 11020424836 & -211585812712 & 4272269155624\\
\hline $\phi_{-2,1}^4 E_4^2$&-16 & 616 & -8512 & 131476 & -1473568 & 21494376 & -500910496 & 10920941572 & -211086335536 & 4272269156968 \\
\hline
\end{tabular}}
\caption{List of $f(m)$ for various seed theories.}
\label{table:fm}
\end{table}

The data illustrates the difference between $M=1$
and higher $M$.
For $M>1$ in the chosen basis, the $f(m)$ grow very rapidly with $m$.
For a generic weak Jacobi form, we expect no significant
cancellations.
Under this assumption we can then estimate the seed theory coefficients from the Cardy formula
\be
c(n, \ell) \sim \exp{\(\pi\sqrt{4Mn - \ell^2}\)}.
\label{eq:ginger}
\ee
In (\ref{fhog}), the dominant contribution to $f(m)$ comes from the term in the sum with maximal polarity. A simple saddle point computation shows this is at
\begin{align}
n_* &= \frac{2m}{M} \nn \\
\ell_* &= 0.
\label{eq:chopsticks}
\end{align}
This gives 
\be
f(m) \sim \exp{(2\pi m)}.
\ee
Note that in Table \ref{table:fm}, this is roughly the growth we see for arbitrary seed theory. This growth of $f(m)$ gives a Hagedorn growth in the density of states.

However, one can envision choosing linear combinations of the basis forms which enjoy cancellations
between the (rapidly growing) $f(m)$ at a given $M > 1$.  We discuss this strategy in the next section,
and find what we conjecture to be the full set of special cases where $f(m)$ grows slowly enough that
the large $N$ symmetric product exhibits supergravity (as opposed to Hagedorn) growth in some regime
beneath the lightest black hole mass.

\subsection{`Very special' weak Jacobi forms}

In this subsection, we define a special class of weight 0 weak Jacobi forms, the `very special'
weak Jacobi forms.  These will be precisely the forms at a given index that (we conjecture) give sub-Hagedorn growth of the polar spectrum, when run through the DMVV formula.

To motivate the definition of `very special' Jacobi forms, let us consider the NS sector elliptic genus with $z=0$, which we can call $Z_{NS}(\tau)$. At index $M$, $Z_{NS}(\tau)$ can be written as a polynomial of degree $M$ in $\kappa(\tau)$, defined in (\ref{eq:california}). Moreover, if $M$ is odd, the polynomial contains only odd powers of $\kappa$; if $M$ is even, only even powers. 

We define a `very special' weak Jacobi form as one which satisfies
\be
Z_{NS}(\tau) = v\kappa(\tau)= vq^{-1/4} + \ldots
\ee
for some positive integer $v$. Note that very special weak Jacobi forms only exist for odd $M$. If there are no cancellations in the genus, at index $M$, then the NS-sector weak Jacobi form starts off as
\be
Z_{NS}(\tau) \sim q^{-M/4} + \ldots.
\ee
Thus to become a very special weak Jacobi form at index $M$, many 
delicate cancellations are needed (from the right-moving Ramond ground states) for the first $\frac{M-1}{2}$ terms. For instance, the elliptic genus of K3, $2\phi_{0,1}$, is a very special weak Jacobi form with $M=1$, and $v=2$; the function $\phi_{0,1}^3 - \phi_{0,1} E_4 \phi_{-2,1}^2$ is very special with $M=3$ and $v=48$.

Now we will derive the analogue of (\ref{fhog}) for seed theories with vanishing ``vacuum" terms. (Recall that in deriving (\ref{fhog}), we assumed that $c(0,-M)$ was nonzero.) Suppose we are considering a seed theory which has the first $a$ terms in $Z_{NS}(\tau)$ vanishing. In other words,
\be
Z_{NS}(\tau) = v q^{\frac{-M+2a}4} + \ldots.
\ee
Let's again define a generating function for the elliptic genus of $\text{Sym}^N(X)$ in the NS sector shifting energy so that the first nonzero term appears at $q^0$. The analog of (\ref{eq:philadelphia}) is then
\begin{align}
\sum_{N=1}^\infty p^N Z^{{\rm Sym}^N(X)}_{NS}(\tau,z) &= \sum_{N=1}^\infty p^N Z^{{\rm Sym}^N(X)}_{EG,R}(q,y\sqrt{q}) y^{NM} q^{N{M-a}/2}\nn\\
&=\prod_{\substack{n\geq1, n \in \mathbb{Z}\\ m\geq0, m \in \mathbb{Z}\\ \ell \in \mathbb{Z}}} ~{1 \over {(1- p^n q^{m+ \ell/2 + Mn/2-an/2}y^{\ell + Mn})^{c(nm,\ell)}}}~.
\label{eq:rice}
\end{align}
Relabelling 
\begin{align}
m' &= m + \frac{\ell}2 + \frac{nM}2 - \frac{an}2 \nn \\
\ell' &= \ell + Mn
\end{align}
and dropping primes, we get
\begin{align}
\sum_{N=1}^{\infty} p^N Z^{\text{Sym}^N(X)}_{NS}(\tau,z) &= \prod_{\substack{n\geq1, n \in \mathbb{Z}\\ m-\frac{\ell}2+\frac{an}2 \in \mathbb{Z}\\ \ell \in \mathbb{Z}}} \frac{1}{(1-p^nq^my^\ell)^{c(nm-\frac{\ell n}2+\frac{an^2}2,\ell-Mn)}}.
\label{eq:sake}
\end{align}
In analogy to (\ref{fhog}), we see that the growth of this, to leading order in $N$ at large $N$, is given by the $q^x$ coefficient of 
\be
\frac{1}{(1-q^m)^{f(m)}}
\ee
where
\be
f(m) = \sum_{n=1}^{\infty} \sum^*_{\ell} c(nm-\frac{\ell n}2+\frac{an^2}2,\ell-Mn).
\label{eq:miso}
\ee
In (\ref{eq:miso}), the * indicates the same limits on $\ell$ as in (\ref{eq:sake}). In particular, if both $a$ and $n$ are odd, $\ell$ is odd for integer $m$ and even for half-integer $m$; if one of $a$ or $m$ is even, $\ell$ is even for integer $m$ and odd for half-integer $m$.

Empirically we find that for very special weak Jacobi forms, $f(m)$ does not grow with $m$, but alternates between two values depending on if $m$ is integral or half-integral. This implies a growth of states as
\be
c_{\Delta} \sim e^{\sqrt{\Delta}}.
\ee In particular, we find for a seed weak Jacobi form of index $M$ that satisfies 
\be
Z_{NS}(\tau) = v \kappa(\tau),
\ee
we get
\be
f(m) = \begin{cases}
\frac{\chi}2 + 8v ~~~ &m \in \mathbb{Z} \\ \frac{\chi}2 + 16v ~~~ &m \in \mathbb{Z} + \half 
\end{cases}
\ee
where $\chi$ is the Witten index of the seed theory. For K3, $v=2$ and $\chi=24$, which
indeed reproduces (\ref{K3f}).

The physical interpretation of the sub-Hagedorn growth of higher index very special weak Jacobi forms is unclear. 
In particular, for these theories, we are counting states $\(\frac{M-1}4\)N$ above the vacuum. The Planck mass is at $\frac{MN}4$ above the vacuum, so at large $M$, we are counting states extremely close to the Planck mass. The fact that all the lower
terms vanish simply means that we do not get a meaningful bound
for the number of such light states.

Note that the left-moving vacuum always leads to a contribution at $q^{-M/4}$.
The only way to eliminate this contribution is if the left-moving
vacuum does not just couple to the right-moving vacuum, but also 
to additional primary fields with differing fermion number. This
implies that the theory has an enhanced symmetry. A trivial example
of this is the case of $T^4$, where the entire elliptic genus
vanishes. This probably simply implies that the standard elliptic
genus is too coarse an index to give any meaningful information
on the number of light states, and that one should consider
a more refined index, such as the one introduced in \cite{Maldacena:1999bp}
for the $T^4$ case.
% It seems likely to us that such such
%a refined index would again show Hagedorn growth if 
%the seed theory has $c>6$. {\rm[DO WE KNOW DIFFERENTLY
%FROM COUNTEREXAMPLES?]}

Empirically we find that for any other non-constant $Z_{NS}(\tau)$, the $f(m)$ grow exponentially, giving Hagedorn growth. (The final case, if $Z_{NS}(\tau)$ is a constant, leads to the symmetric product elliptic genus having only one non-vanishing term, corresponding to taking $N$ copies of the non-vanishing term, and thus does not have interesting growth.)
In the next subsection we will take some steps towards proving this.

\subsection{Hecke operators}
To prove this conjecture, it seems natural to
use the language of Hecke operators and start
from (\ref{DMVVHecke}).
First consider the case $M=1$. 
The specialization $\chi_L$ of the $L^{\text{th}}$ Hecke transform of
the unique weak Jacobi form $\phi_{0,1}$ is given by
\be
\chi_L(q)= \frac{1}{L}q^{-{L/4}}+ \sum_{ad=L}\chi^0(d) + o(q)\ .
\ee
where
\be
\chi^0(d) = \begin{cases}
a^{-1}c(0) = \frac{10}a &~~~d \equiv 0 ~(\text{mod}~ 4)\\
a^{-1}2c(-1) = \frac{2}a &~~~d \equiv 2 ~(\text{mod}~ 4)\\
0 &~~~\text{else}
\end{cases}
\ee
as is shown in Appendix~\ref{app:HeckeK3}.

For $L$ even/odd, we can write 
\be
\chi_L(q) = \sum_{n=-L/4}^\infty d(n) q^n
\ee
 as an even/odd polynomial in $\kappa$.
Using (\ref{kappatrafo}), we can obtain
the asymptotic behavior by repeating the Cardy argument. 
This gives
\be
d_L(n) \sim (-1)^L \left(16n^3L^3\right)^{-1/4}e^{2\pi\sqrt{Ln}}\ ,
\ee
where the sign $(-1)^L$ comes from the
behavior of $\kappa$ under the $S$ transform.
The generating function for the symmetric orbifold
elliptic genus is
\be
\mathcal{Z} = \exp \sum_L p^L T_L\phi\ .
\ee
For the shifted NS partition function at $N=\infty$ we
thus have
\be
\chi_\infty= \exp \sum_L (q^{L/4}\chi_L(q)-1/L)\ ,
\ee
so that that the $n^{\text{th}}$ term of $\log \chi_\infty(q)$ is
\be\label{dnsum}
d(n) = \sum_{L=1}^{4n} d_L(n-L/4)\ .
\ee
The biggest term in (\ref{dnsum}) is thus
\be
d_{2n}(n)\sim 2n^{-3/2}e^{2\pi n}\ .
\ee
If there were no cancellations, we would
thus have Hagedorn growth already in the
exponent, so that $\chi_\infty(q)$ would have
Hagedorn growth too. Crucially, because of the factor $(-1)^L$,
there are a lot of cancellations. In fact
\be
\log \chi_\infty(q) = -22 \sqrt{q}+25 q -\frac{1}{3} 88 q^{3/2}+\frac{53 q^2}{2}-\frac{132 q^{5/2}}{5}+\ldots
\ee
so that the coefficients grow very slowly. This explains
why we do not get Hagedorn growth in this case.

For $M=2$, the Hecke transforms
give
\be
\chi_L(q) = \frac{1}{L}q^{-L/2} +\mc O(1)\  .
\ee
Note that $\chi_L$ for the different weak Jacobi forms differ only
in the constant term.
Also note that now for every $L$ we get an even
polynomial in $\kappa$, which means that
all coefficients $d_L(n)$ are positive. This means
that there are no cancellations, so that we indeed
get Hagedorn growth. 

%
%A similar argument should carry over to even $M$:
%The leading term $q^{ML/4}$ fixes the sign of the coefficients,
%since $\chi_L$ is invariant under $S$, which means that
%there cannot be any cancellations between different $L$s.
%The growth is thus necessarily Hagedorn. 
%If $\chi_L(q)= L^{-1}q^{-LM/4}+\ldots$, then we have
%\be
%d_L(n) \sim (-1)^{LM}M\left(\frac{M}{16n^3L^3}\right)^{1/4}
%e^{2\pi\sqrt{MLn}}
%\ee

We believe that there should be a similar argument that
establishes Hagedorn growth for all weak Jacobi forms
that are not very special.

\section{Some simple wreath products}
\label{sec:wasabi}
\subsection{Symmetric orbifolds of families of CFTs}
The results of our previous section suggest 
that the only symmetric orbifolds that can
have supergravity growth come from seed theories
with $c=6$. To find other examples, 
let us therefore generalize the
discussion to more general permutation orbifolds.

In particular, consider ${\rm Sym}^a (\C_b)$ with $a$ fixed and $b=1,2,3,\ldots$ indexing the sequence of CFTs $\C_b$ asymptoting to infinite central charge.  
Eventually we will be interested in the elliptic genus of ${\rm Sym}^a({\rm Sym}^b(K3))$,
but let us first discuss some general points.

Let us start by discussing partition functions rather than
elliptic genera. This has the advantage that no cancellations
can occur. We will see that that behavior of the two is very similar.
Take a partition function
$Z_b$ of central charge $b$. If we do not make any
additional assumptions, the symmetric orbifold Sym$^a(Z_b)$
will have the following behavior:
\be
\rho_{a,b}(\Delta) \sim
\begin{cases}
e^{2\pi\sqrt{ab\Delta/6}} ~~~& \Delta > ab/6\\
e^{2\pi \Delta} ~~~& ab/6>\Delta>b\\
\text{non-universal} ~~~& b>\Delta
\end{cases}
\ee
To say something about the region $\Delta<b$, we
need to make additional assumptions on $Z_b$.
Assume that the growth of $Z_b$ is governed
by an effective central charge $\hat c$ such that
\be\label{ceff}
\rho_b(\Delta) \sim e^{2\pi\sqrt{\hat c \Delta/6}}\ .
\ee
Examples of such families are for instance
the putative extremal CFTs or indeed the holographic
dual to the D1-D5 system at the supergravity point.
First note that the twisted states of the $\text{Sym}^a$
orbifold have weight $\Delta_{tw} > b/16$ and thus
do not contribute here. The number of states in
the untwisted sector coming from $K$ non-trivial
states and $a-K$ vacuum states is
\be
\sim\frac{e^{2\pi\sqrt{\hat c \Delta K/6}}}{K!}\sim 
e^{2\pi{\sqrt{\hat c \Delta K/6}}-K\log K +K}
\ee
This is maximized for 
$K\sim \frac{\pi^2\hat c \Delta/6}{(\log\pi^2\hat c\Delta/6)^2}$
and gives 
\be
\rho_{ab}\sim \exp\left(
\frac{\pi^2\hat c \Delta/6}{\log\pi^2\hat c\Delta/6}\right)\ .
\ee
This growth is sub-Hagedorn, but it is much faster
than supergravity.
Note that $K<a$. If $a<b$, we roughly have
\be
\rho_{ab}(\Delta) \sim 
\begin{cases}
e^{2\pi\sqrt{\hat c a\Delta/6}} ~~~& b>\Delta > a\\
e^\frac{\pi^2\hat c \Delta/6}{\log\pi^2\hat c\Delta/6} ~~~& a>\Delta
\end{cases}
\ee
This suggests that there is a new phase
in the regime $a>\Delta$. Depending whether
$a<b$, there is also a supergravity phase for
$a<\Delta<b$.

\subsection{K3-based theories}

Consider a theory where the seed (NS-sector) elliptic genus has the form
\begin{equation}
Z_{NS}(q) = \sum_{n=0}^{\infty} c(n) q^{n - b/4}
\end{equation}
where
\begin{equation}
\label{wombat}
c(n)=
\begin{cases}
e^{\sqrt{n}}~~~&n < b/4\\
e^{\sqrt {bn}}~~~&n > b/4
\end{cases}
\end{equation}
This is, for instance, the case for ${\rm Sym}^b(K3)$ at large $b$.

Now, consider ${\rm Sym}^a({\rm Sym}^b(K3))$, with $a$ fixed and taking $b$ large. The value of $a$ labels which family of CFTs we are considering, and $b$ is taken to infinity in each family. The key point is that by introducing a new parameter $a$, we hope to introduce a new scale in the problem. With this in mind, we estimate the growth of states in four different regimes: states with dimension $\Delta \ll a$, $a\ll\Delta \ll b$, $b \ll \Delta \ll ab$, and $ab \ll \Delta$.

For states with $\Delta \ll a$, we of course cannot excite states in all $a$ copies of the seed theory. Suppose we excite states in some $k$ copies to the $\frac{\Delta}{k}$ level. This would give us a degeneracy of roughly
\be
\sim \frac{\(e^{\sqrt{\frac{\Delta}k}}\)^k}{k!} \sim \frac{e^{\sqrt{\Delta k}}}{k!} \sim e^{\sqrt{\Delta k}-k \log{k} + k}.
\ee
This is maximized for $k \sim \frac{\Delta}{(\log \Delta)^2}$ and gives
\begin{equation}
c_{{\rm Sym}^a({\rm Sym}^b)}(\Delta) \sim {\rm exp}\(\frac{\Delta}{\log \Delta}\)~.
\end{equation}
%\subsubsection{$\Delta \ll a$}

%\subsubsection{$a \ll \Delta \ll b$}
In the regime of states with
$a \ll \Delta \ll b$, one only draws states (to symmetrize) from the upper case in (\ref{wombat}).  The leading
growth is easily estimated via saddle point (roughly coming from taking a contribution to $\Delta$ of ${\cal O}(\Delta/a)$
from the $a$ copies of ${\rm Sym}^b(K3)$), and gives growth
\begin{equation}
c_{{\rm Sym}^a({\rm Sym}^b)}(\Delta) \sim {\rm exp}(\sqrt{a\Delta})~.
\end{equation}

In the regime where $b \ll \Delta \ll ab$, we can suddenly draw from black hole states in the seed theory in (\ref{wombat}). Taking states with energy $\mc{O}(b)$ from $\frac{\Delta}{b}$ copies of the seed gives a growth
\begin{equation}
c_{{\rm Sym}^a({\rm Sym}^b)}(\Delta) \sim {\rm exp}(\Delta)~.
\end{equation}
This shows a Hagedorn density of states, which hints that stringy states show up in the dual at this energy. 

Finally in the regime where $ab \ll \Delta$, we are in the Cardy regime of the full theory, which means
\begin{equation}
c_{{\rm Sym}^a({\rm Sym}^b)}(\Delta) \sim {\rm exp}(\sqrt{a b\Delta})~.
\end{equation}

Thus in total the growth goes as
\begin{align}
\rho(\Delta)=
\begin{cases}
\exp{(\frac{\Delta}{\log{\Delta}})} ~~ &\Delta\ll a  \\
\exp{(\sqrt{a\Delta})} ~~ &a\ll \Delta\ll b  \\
\exp{(\Delta)} ~~ &b \ll \Delta\ll ab  \\
\exp{(\sqrt{ab\Delta})} ~~ &ab\ll \Delta.
\end{cases}
\label{eq:avocado}
\end{align}

In our derivation of the first three lines of (\ref{eq:avocado}), we have only considered states in the untwisted sector of the Sym$^a$ orbifold. However, states in the twisted sector of the Sym$^a$ orbifold will start contributing at $\Delta\sim\mc{O}(b)$, so the neglect of the twisted sector states in the first two lines of (\ref{eq:avocado}) is justified. Moreover, the partition function in the regime $b \ll \Delta \ll ab$ grows exponentially, as seen in e.g. \cite{BKM, Keller}, which puts an upper bound on the growth of the genus. Combining this with the lower bound from the untwisted sector, we see that the third line in (\ref{eq:avocado}) is also accurate. The fourth line comes from Cardy formula, which of course takes into account the twisted sectors.

\begin{figure}[h!]
\centering
  \includegraphics[width=0.8\textwidth]{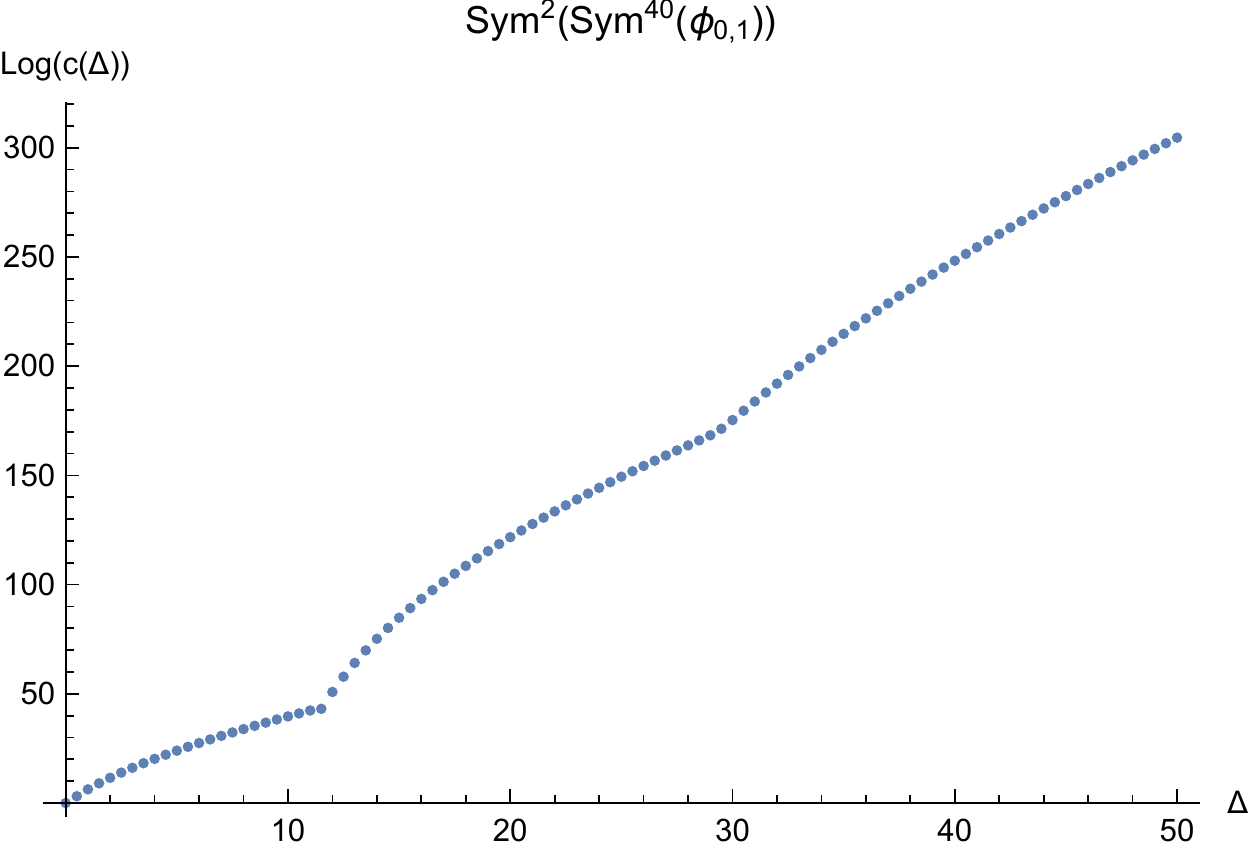}
  \caption{Growth of coefficients in NS sector of $\text{Sym}^2(\text{Sym}^{40}(\phi_{0,1}))$. Note the three very distinctive regions of $a < \Delta < \frac{b}4$, $\frac{b}4 < \Delta < \frac{ab}4$, and $\frac{ab}4<\Delta$. (The region $\Delta < a$ is too small here to notice.)}
  \label{fig:tobiko}
\end{figure}

In Figure \ref{fig:tobiko}, we show a plot of the growth of states of $\text{Sym}^2(\text{Sym}^{40}(\phi_{0,1}))$, where we see evidence of (\ref{eq:avocado}) in the different regimes\footnote{We compute this for $\phi_{0,1}$ instead of the elliptic genus of K3; these differ by a factor of 2. This is simply to avoid the large Ramond ground state degeneracy of the K3 case, which shows up in the elliptic genus.  One can achieve $\phi_{0,1}$ as the elliptic genus of a known manifold -- the Enriques surface \cite{Eguchi}. }.
This is all compatible with our initial observation about the existence of moduli
in the twisted sector: ${\rm Sym}^b(K3)$ does have moduli with which
we can deform away from the orbifold point. In this way we can
move to a point in the moduli space where the growth of states in the full
partition function is
as in (\ref{ceff}). By the previous analysis, ${\rm Sym}^a$ of this will then indeed 
be as in (\ref{eq:avocado}).

%\subsubsection{$b \ll \Delta \ll ab$}

%\subsubsection{$ab\ll\Delta$}

\section{Discussion}
\label{sec:jiro}

The main focus of this paper was to provide examples where the elliptic genus, a 2d supersymmetric
index, could be a useful indicator of an emergent macroscopic space-time geometry that is computable directly in quantum field theory.  The physical criteria which slow-growth of the
genus indicate, include a sparse spectrum of low-energy modes, or equivalently (via the modularity arguments in \cite{HKS,BCKMP}) a good match to Bekenstein-Hawking entropies of BPS black holes.  
Another equivalent criterion (though not manifestly so) requires the existence of a conventional Hawking-Page phase transition.  The main advantages of the genus relative to the full partition function 
are that it is computable, and (subject to a non-cancellation hypothesis) it captures the growth characteristic of the sparsest regions in moduli space.  It thereby serves immediately as an indicator of whether ${\it any}$ point in the CFT moduli space can be dual to weakly curved gravity.

There are other natural candidates to serve as order parameters for emergent spacetime.
The entanglement entropy at large central charge enjoys various special properties
\cite{Hartman}, and matching
entanglement computed via the Ryu-Takayanagi formula in suitable geometries to the entanglement
in a given quantum field theory could serve as a possible test.  More recently, it has been proposed
that CFTs whose out-of-time-order four point functions saturate a certain `chaos bound' may be 
dual to Einstein gravity \cite{Steve}.   It is interesting to consider the precise relationship between
these criteria as indicators of emergent space-time geometry; this is the subject of ongoing 
research \cite{toappear}.

Even within the world of 2d supersymmetric CFTs and density of states bounds, much work remains to be done.  We focused here on theories with $(2,2)$ supersymmetry.  While our work could be easily generalized to classes of theories with less supersymmetry, a wide class of $(0,4)$ theories which
arise on MSW black strings \cite{MSW} -- the strings arising as M5-branes wrapping divisors in Calabi-Yau threefold compactifications of M-theory -- do not fall into this category.  For these theories, the
conventional elliptic genus is uninteresting, and one should compute instead an improved object,
the M5-brane elliptic genus of \cite{GaiottoYin}.  This object is still modular, but has more involved
properties than a weak Jacobi form.  Even for a small number of M5-branes, the computations
involved become quite sophisticated \cite{Klemm}.  It would be interesting to develop a generalization of the criteria derived in \cite{BCKMP} and explored in this paper, to that class of theories.

A related issue is the existence of theories with enhanced symmetry, where the elliptic genus
fails to capture the BPS information, but a refined index can do so \cite{Maldacena:1999bp,refined}.
It seems likely that by suitable refinement, one can define an improved index that could capture more information about the spectra
of the theories yielding `very special' Jacobi forms as their elliptic genera.  

Finally, we note that the modular properties of the elliptic genus have been crucial in all of our considerations.  However, it is clear that the individual terms in the $q,y$ expansion are themselves
constants across the moduli space.  So, for instance, instead of considering the NS elliptic genus
at $y=1$, one could gain more information by taking a fixed power of $q$ in the NS genus and
summing the absolute values of the coefficients of the various powers of $y$ which appear.
The resulting counting function would certainly not have any elegant modular properties.  But 
numerical studies of this function could yield more information about the detailed growth of 
the spectrum in various classes of theories (for instance, those where the coefficients vanish
after specialization to $y=1$).

\bigskip
\centerline{\bf{Acknowledgements}}
\bigskip

We thank A. Belin, M. Cheng, J. de Boer, E. Dyer, A. L. Fitzpatrick, A. Maloney, G. Moore, and E. Perlmutter for very helpful discussions about related matters.
We are grateful to the Perimeter Institute workshop on `(Mock) modular forms, moonshine and 
string theory,' the LMS-Durham workshop on `New moonshines, mock modular forms and string theory,' and the Aspen Center for Physics for hospitality while this work was in progress. We especially thank the juicers at PI.  N.B. is supported by a National Science Foundation Graduate Fellowship and a Stanford Graduate Fellowship. N.M.P. is supported by a National Science Foundation Graduate Fellowship, and also gratefully acknowledges the University of Amsterdam for hospitality and the Delta Institute for Theoretical Physics for additional support while this work was being completed. S.K. acknowledges the support of the National Science Foundation via
grant PHY-1316699. C.A.K. is supported by the 
Swiss National Science Foundation through the NCCR SwissMAP.

\appendix

\section{Derivation of K3 Growth}
\label{app:soysauce}

In this appendix, we will derive equation (\ref{K3f}), namely that for seed theory K3, 
\begin{align}
f(m) = \sum_{n=1}^{\infty} \sum^{\ell=2m}_{\substack{\ell=-2m\\\ell\equiv2m\text{~(mod~}2)}} c(nm-\frac{n\ell}2,\ell-n)=\begin{cases}
28 ~~ m\in\mathbb{Z}  \\
44 ~~ m\in\mathbb{Z}+\frac12.
\end{cases}
\label{eq:2844}
\end{align}

First note that for K3, from spectral flow $c(n,\ell) = c(4n-\ell^2)$. Thus we can rewrite (\ref{eq:2844}) as
\be
f(m) = \sum^{\ell=2m}_{\substack{\ell=-2m\\\ell\equiv2m\text{~(mod~}2)}}  \sum_{n=1}^{\infty} c(4mn-n^2-\ell^2).
\ee
Define $\tilde{n} = n - 2m$ to get
\begin{align}
f(m) &= \sum^{\ell=2m}_{\substack{\ell=-2m\\\ell\equiv2m\text{~(mod~}2)}}  \sum_{\tilde{n}=-2m+1}^{\infty} c(4m^2-\tilde{n}^2-\ell^2)\nn\\
&= \sum^{\ell=2m}_{\substack{\ell=-2m\\\ell\equiv2m\text{~(mod~}2)}} \( \sum_{\tilde n=-\infty}^{\infty} c(4m^2-\tilde n^2-\ell^2) \) -  c(-\ell^2)  \nn \\
&= \sum^{\ell=2m}_{\substack{\ell=-2m\\\ell\equiv2m\text{~(mod~}2)}} \( \sum_{\tilde n=-\infty}^{\infty} c\(m^2-\frac{\ell^2}4, \tilde n\) \) - c(0,\ell). 
\label{eq:tuna}
\end{align}
The first term in (\ref{eq:tuna}) vanishes unless $m^2-\frac{\ell^2}{4}=0$ (this is the statement that Witten index has no $q$-dependence). These are precisely the $\ell = \pm 2m$ terms in the sum, which each contribute $\chi_{K3}=24$. The second term, if $m \in \bb Z$, is $c(0,0)=20$; if $m \in \bb Z + \half$, is $c(0,1)+c(0,-1)=4$. Thus we reproduce (\ref{eq:2844}).

\section{Hecke transforms of K3}\label{app:HeckeK3}
Let us gather the contributions to the polar 
and the constant terms
of $\chi$. 
Define the polarity as $A:= m+\ell/2+d/4$ and $d=4m+d'$ such that
$A = d/2+\ell/2-d'/4$. 
We know that $c(m,\ell)$ vanishes if
\be
4Mm-\ell^2 < -M^2\ .
\ee
The condition on $\ell$ is thus
$\ell^2\leq d^2 - d'd+1$.
We can then estimate
\be
A \geq  \frac{1}{2}\frac{d'^2/4-1}{d-d'/2+\sqrt{d^2-d'd+1}}\ .
\ee
With $d'\leq d$ it follows automatically that polar terms can
only come from $|d'|<2$, and constant terms from $|d'|\leq2$.
Analyzing case by case we find that the only
polar term comes from $d'=1, d=1, a=L$ with $m=0$ and $\ell=-1$.
Define $\chi_L(q)$ as the spectrally flowed and
specialized Hecke transform of $\phi$.
We thus have
\be
\chi_L(q) = \frac{1}{L}q^{-L/4} + \mc O(1)\ .
\ee
The constant terms come from $d'=-2,0,2$. 
For $d'=0$, we have a contribution from $m=d/4, \ell=-d$.
For $d'=\pm2$, we get a contribution from $m=(d\mp2)/4,\ell = -d\pm1$.
Since $m$ has to be integer, in total we get
\be
\chi^0(d) = \begin{cases}
a^{-1}c(0) = \frac{10}a &~~~d \equiv 0 ~(\text{mod}~ 4)\\
a^{-1}2c(-1) = \frac{2}a &~~~d \equiv 2 ~(\text{mod}~ 4)\\
0 &~~~\text{else}
\end{cases}
\ee
The total result is thus
\be
\chi_L(q)= \frac{1}{L}q^{-{L/4}}+ \sum_{ad=L}\chi^0(d) + o(q)\ .
\ee

\end{document}